\documentclass[onecolumn]{aa}
\usepackage{graphicx}
\usepackage{txfonts}

\begin{document}
   \title{Naked active galactic nuclei}

   \author{M.R.S. Hawkins
          \inst{1}
          }

   \offprints{M.R.S. Hawkins}

   \institute{Institute for Astronomy (IfA), University of Edinburgh,
              Royal Observatory, Blackford Hill, Edinburgh EH9 3HJ\\
              \email{mrsh@roe.ac.uk}
             }

   \date{Received April 21, 2004; accepted May 29, 2004}

   \abstract{
In this paper we report the discovery of a new class of active
galactic nucleus in which although the nucleus is viewed directly,
no broad emission lines are present.  The results are based on a survey
for AGN in which a sample of about 800 quasars and emission line
galaxies were monitored yearly for 25 years.  Among the emission line
galaxies was the expected population of Seyfert 2 galaxies with only
narrow forbidden
lines in emission, and no broad lines.  However, from the long term
monitoring programme it was clear that some 10\% of these were strongly
variable with strong continuum emission.  It is argued that these
objects can only be Seyfert 1 galaxies in which the nucleus is viewed
directly, but in which broad emission lines are completely absent.
We compare these observations with other cases from the literature
where the broad line region is reported to be weak or variable, and
investigate the possibility that the absence of the broad line
component is due to reddening.  We conclude that this does not account
for the observations, and that in these AGN the broad line region is
absent.  We also
tentatively identify more luminous quasars from our sample where the
broad emission lines also appear to be absent. The consequences of this
for AGN models are discussed, and a case is made that we are seeing AGN
in a transition stage between the fuel supply from a surrounding star
cluster being cut off, and the nucleus becoming dormant.
   }

   \maketitle
%

\section{Introduction}

The diverse phenonema which characterise the class of objects
known as Active Galactic Nuclei (AGN) have been reconciled in a unified
scheme (\cite{a93}) in which viewing angle plays a crucial role in the
observed properties of the system.  An important feature of the
unified model is the distinction between Seyfert 1 galaxies and
quasars where the active nucleus is viewed directly, along with
emission from fast moving gas clouds surrounding it, and Seyfert 2
galaxies where this central region is hidden from view by a dusty
torus.  Although there is some difference in the observed velocity
structure in the clouds, they are taken to be a fundamental feature
of AGN.

From an observational point of view, there is a clear distinction
between Seyfert 1 and 2 galaxies. The spectra of Seyfert 1 galaxies
show broad emission lines with a velocity dispersion of around 6000
to 10,000 km sec$^{-1}$ together with narrow relatvely weak forbidden
and permitted lines.  The narrow lines typically have a velocity
dispersion of less than 1000 km sec$^{-1}$, and a
[O III]\,$\lambda 5007/$H$\beta$ ratio
around 10.  In addition, Seyfert 1 galaxies have a strong variable
component of continuum emission from the active nucleus, whereas for
Seyfert 2 galaxies where the nucleus is obscured, only a faint
reflection of the nucleus is seen in polarised light (\cite{c00}).
Most Seyfert galaxies fit comfortably into one or other of these
classes, consistent with predictions from the unified scheme.

The unified scheme incorporates the idea that all AGN have a broad
line region (BLR) which would be visible but for the obscuring torus.
However, the place of the broad line region in the evolution of AGN
is far from clear, and the possibility that some AGN may not contain
a BLR has not been ruled out.  There are a number of interesting
departures from the norm, the most prominent of which is a class
of `narrow line' Seyfert 1 galaxies (\cite{o85,r00}).  In these AGN the
emission lines are less broad than normal, in extreme cases with a
width close to the forbidden lines from the narrow line region, but
they are always strong, with a [O III]\,$\lambda 5007/$H$\beta$ ratio
less than 3.  There are also other situations where emission lines 
are very weak.  For example there are a number of quasars where the
Lyman-$\alpha$ line appears to be cut away to a thin spike by
absorption, and in one notable case (\cite{f99}) appears to be absent
altogether.  Another interesting example (\cite{m95}) is PG 1407+265
with H$\beta$ and Lyman-$\alpha$ very weak, but a strong broad
H$\alpha$ line.  Emission line strengths are also known to vary
strongly in some Seyfert 1 galaxies, apparently in response to
changes in the brightness of the continuum.  NGC 4151 provides a good
example (\cite{c87}), but the broad lines by no means disappear, the
C IV\,$\lambda 1550$ line changing in width between 2600 and 5500
km s$^{-1}$ as the continuum brightness varies.  Lastly it is worth
mentioning the long sought class of type 2 quasar.  These are luminous
counterparts of Seyfert 2 galaxies, and were expected to exist on the
basis of the unified model.  Some good examples are now known
(\cite{n02}), and apart from their luminosity (at $M_{R} \sim -23$)
they have properties closely resembling Seyfert 2 galaxies.  They have
virtually no continuum emission, and most of their observed flux comes
from strong narrow forbidden lines.  It is clear that in these objects
we are not looking directly at the nucleus, which is obscured by a
dusty torus as for Seyfert 2 galaxies.

In this paper we report the discovery of a new class of AGN where the
high velocity clouds are completely absent, although the nucleus is
viewed directly, unobscured by the torus.  These naked AGN have the
spectral properties of Seyfert 2 galaxies, with the exception that
they have a strong variable continuum implying that we are seeing the
active nucleus directly.  This result is discussed in the context of
current AGN models.

\section{The Seyfert galaxy survey}

\subsection{Sample selection}

Over the last 25 years a large-scale monitoring programme for AGN
has been carried out with a view to characterising the nature of the
optical variations and putting constraints on the structure of the
accretion disc (\cite{h96,h00,h02,h03}).  The survey is based on a long
series of photographic plates from the UK 1.2m Schmidt telescope in
Australia, measured with the COSMOS and SuperCOSMOS measuring machines
at Edinburgh.  The survey area comprises the central 20 deg$^{2}$ of
the ESO/SERC field 287 centred on 21h 28m, -45$^{\circ}$ (1950) and
plates were taken every year from 1977 to 2002 in the $B_{J}$ passband
(IIIa-J emulsion with a GG395 filter).  Extensive coverage was also
obtained in other passbands allowing AGN to be selected on the basis
of colour and variability (\cite{h00}).

The measurement of the photographic plates provided a catalogue of
some 200,000 objects in the 20 deg$^{2}$ area, together with UBVRI
colours.  There were also yearly measurements in $B_{J}$ going back to
1977.  Candidate AGN for follow-up were selected from this catalogue
on the basis of their blue colour ($(U-B) < 0$) or variability
according to the criteria given in Hawkins (1996).  This effectively
required that the amplitude of variation exceeded 0.4 magnitudes.
In all, around 1500 candidates down to a magnituide limit of
$B_{J} < 21.5$ were selected as AGN candidates for follow-up
spectroscopic study.

\subsection{Spectroscopy}

\begin{figure*}
\setlength{\unitlength}{1mm}
\begin{picture} (200,235) (-10,0)
\includegraphics[width=0.9\textwidth]{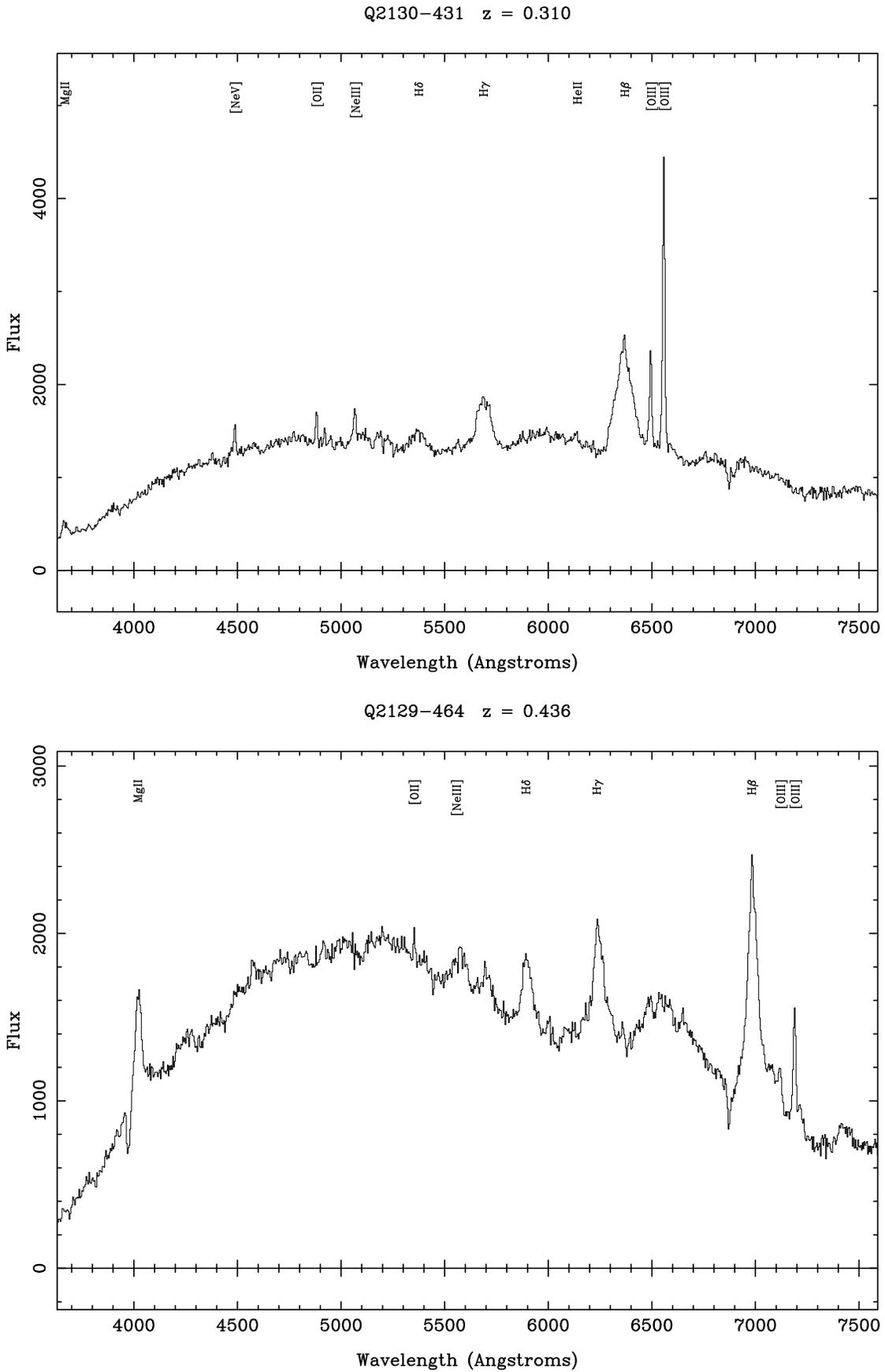}
\end{picture}
\caption{Spectra of typical Seyfert 1 galaxies (flux in arbitrary
 units).
 \label{fig1}}
\end{figure*}

\begin{figure*}
\setlength{\unitlength}{1mm}
\begin{picture} (200,235) (-10,0)
\includegraphics[width=0.9\textwidth]{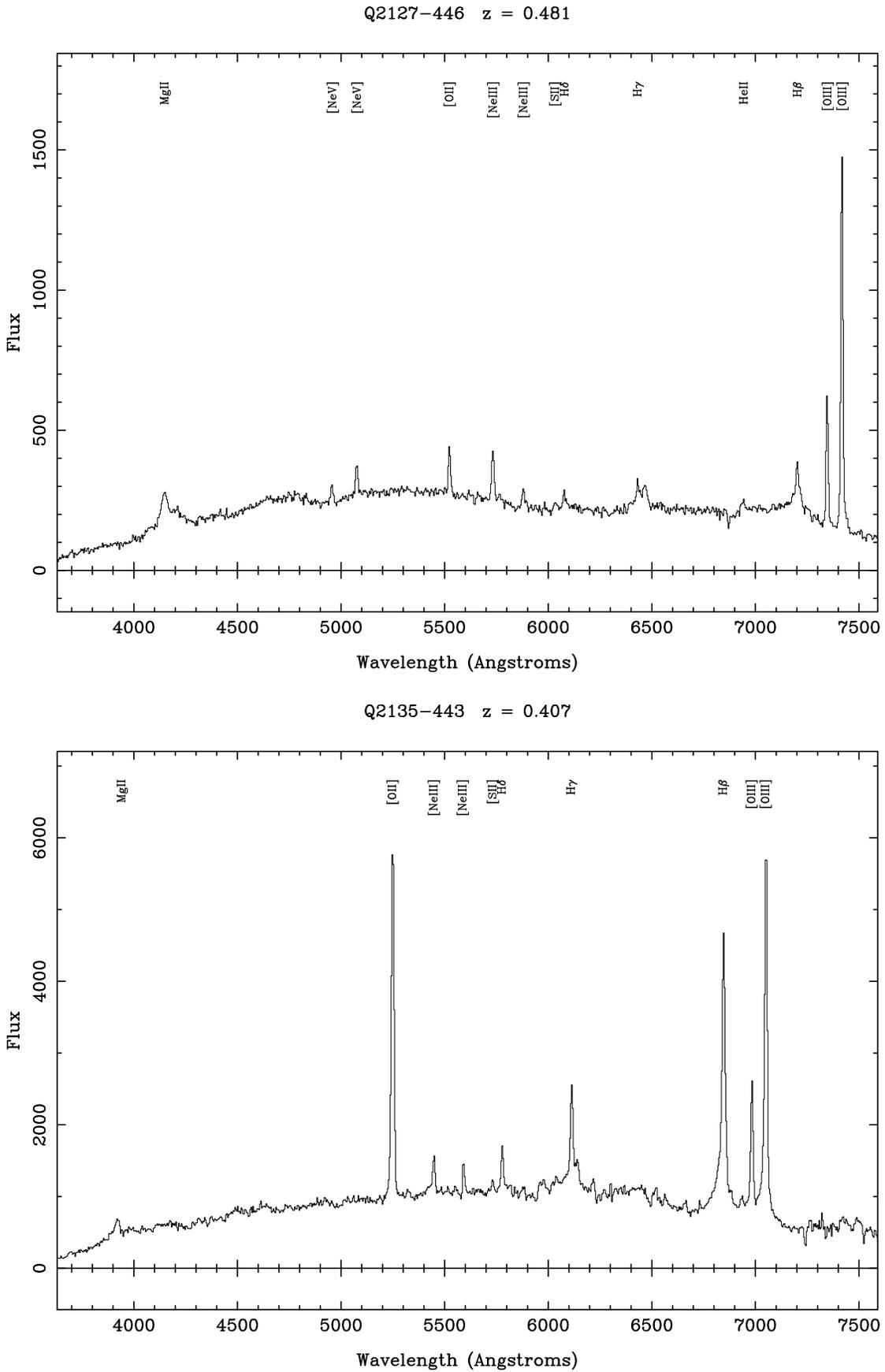}
\end{picture}
\caption{Spectra of typical narrow line Seyfert 1 galaxy (top) and LINER
 (bottom) (flux in arbitrary units).
 \label{fig2}}
\end{figure*}

\begin{figure*}
\setlength{\unitlength}{1mm}
\begin{picture} (200,235) (-10,0)
\includegraphics[width=0.9\textwidth]{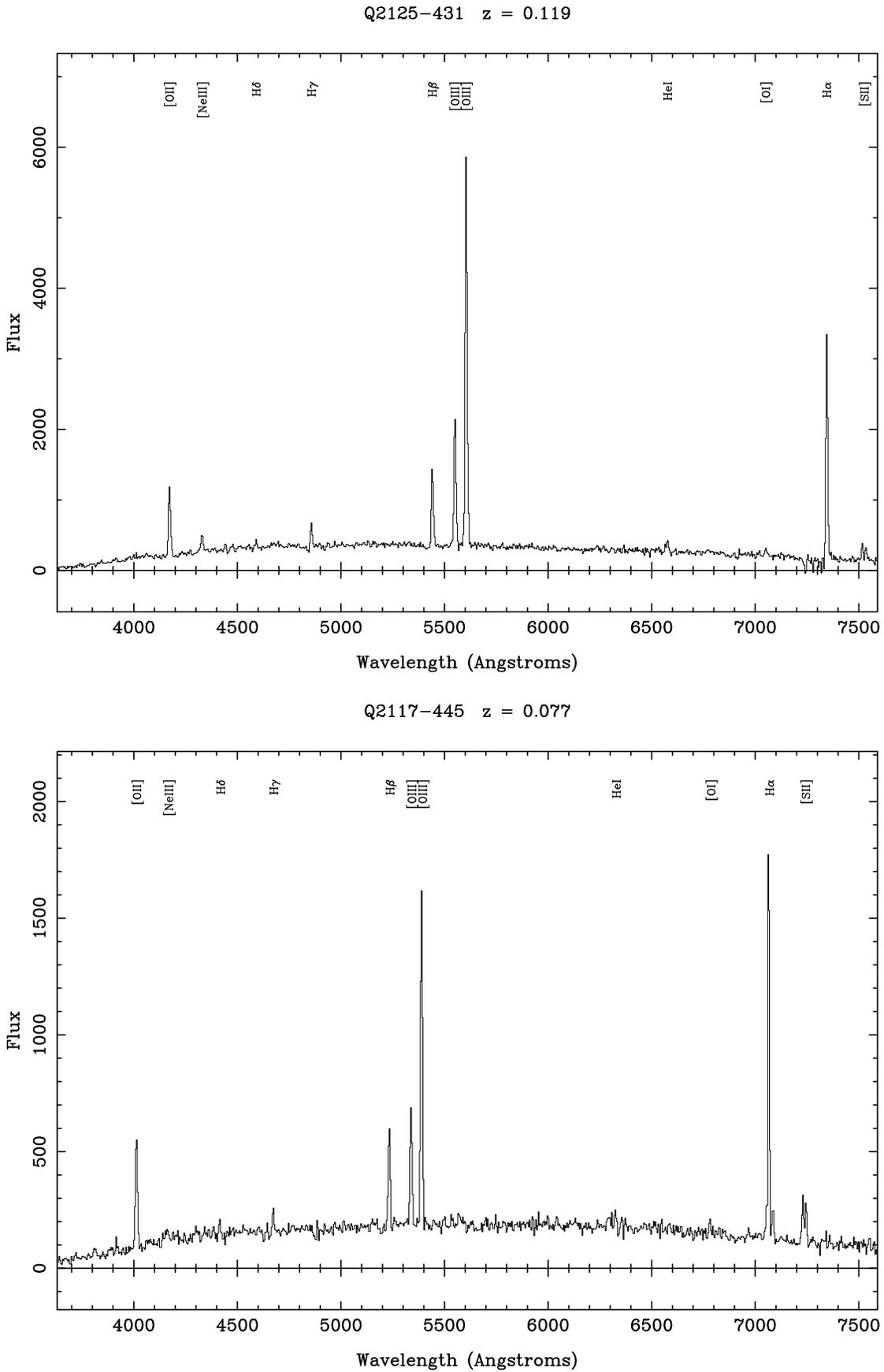}
\end{picture}
\caption{Spectra of typical Seyfert 2 or starburst galaxies (flux in
 arbitrary units).
 \label{fig3}}
\end{figure*}

\begin{table*}
\caption{Seyfert galaxy line widths (km s$^{-1}$)}
\begin{tabular}{rrrrrrrrrrrrrrr}
\multicolumn{3}{c}{RA (1950)}&\multicolumn{3}{c}{Dec (1950)}&
 z&$\delta B$&[O II]&[Ne III]&H$\beta$&[O III]&[O III]&H$\alpha$&O/H\\
&&&&&&&&&&&&&&\\
21&17&36.57&-44&36&49.0&0.391&0.71&1117&1058&1802& 892& 850&   0& 2.89\\
21&17&53.12&-44&32& 6.9&0.077&0.24& 972&   0& 670& 714& 643& 471& 3.23\\
21&17&55.10&-44&46&33.3&0.175&0.19& 943&   0& 691& 687& 626& 378& 2.86\\
21&18&21.53&-44&44&25.7&0.365&0.50& 989& 765&3246& 792& 717&   0& 2.29\\
21&18&24.15&-45&11&18.1&0.272&0.29& 899& 813& 642& 557& 557&   0& 4.64\\
21&19&48.47&-43&57& 2.2&0.092&0.24& 927& 745& 699& 627& 550& 369& 4.01\\
21&20&40.52&-45&48&52.1&0.128&0.29&1202&   0& 761& 704& 614& 508& 3.17\\
21&22&49.22&-44&24&19.4&0.311&0.92& 790&   0& 351& 714& 699&   0& 5.89\\
21&23&31.47&-42&56&35.7&0.141&1.26& 973&   0& 553& 301& 696& 429& 1.55\\
21&23&34.42&-43&38&39.1&0.469&0.19& 983&   0& 738& 998& 531&   0& 0.81\\
21&23&46.41&-42&55&54.5&0.250&0.43& 860&   0& 560& 528& 531&   0& 3.98\\
21&24&18.01&-43&49&22.9&0.294&0.24& 863& 468& 657& 563& 555&   0& 1.45\\
21&24&22.46&-42&59&37.9&0.179&0.16& 976&   0& 628& 497& 696&   0& 4.41\\
21&24&24.99&-42&51&51.5&0.099&0.61& 851&   0&1049& 862& 703& 456& 1.67\\
21&24&54.69&-45& 4&54.4&0.388&0.32& 989&   0& 584& 629& 680&   0& 2.09\\
21&25&50.21&-43&39& 8.0&0.252&0.19& 889& 981& 612& 661& 572&   0& 2.89\\
21&25&54.29&-43& 8&56.7&0.119&0.27& 986&1103& 721& 623& 666& 452& 4.53\\
21&26&10.02&-42&56&30.8&0.410&0.74&1034&1375&3819&1040&1052&   0& 0.55\\
21&26&30.00&-42&47&44.8&0.291&0.43& 982&   0& 694& 682& 652&   0& 2.55\\
21&26&48.98&-43&44&12.9&0.481&0.76&1086& 985& 834& 786& 772&   0&10.83\\
21&26&51.91&-42&46& 0.8&0.333&0.59& 884&   0&2040& 815& 782&   0& 1.31\\
21&26&52.50&-44&23&42.9&0.230&0.25& 880&1070& 590& 524& 494&   0& 3.97\\
21&27&10.40&-42&52&58.3&0.158&0.14& 804& 920& 605& 609& 600&   0& 2.65\\
21&27&22.15&-43&54& 7.8&0.188&0.20&1002&1698& 674& 689& 660& 475& 3.82\\
21&27&28.22&-43&20& 3.5&0.017&0.19&1090&   0& 673& 664& 666& 502& 3.14\\
21&27&33.98&-44&39&18.9&0.481&1.14&1012&1062&1676& 701& 746&   0& 4.65\\
21&28&24.69&-45&37& 3.4&0.202&0.74&1133&   0&3744& 670& 558&1995& 0.40\\
21&29& 5.73&-46&24&19.9&0.140&0.15& 746&1822& 575& 584& 594& 446& 2.16\\
21&29& 9.29&-44&15& 7.9&0.267&0.56&1341&1264&6239&1025&1116&   0& 2.28\\
21&29&19.60&-42&48& 8.8&0.193&0.79&   0&   0&2395& 637& 595&1911& 0.41\\
21&29&58.80&-42&49&26.0&0.194&0.23& 884& 803& 541& 557& 508& 463& 4.35\\
21&30& 4.24&-43& 7&44.2&0.266&1.20& 677& 906& 426& 652& 661&   0&16.36\\
21&30& 7.17&-43&25&42.9&0.456&1.17&1516&   0&1637&1732&1893&   0&20.82\\
21&30&34.23&-46&36& 1.2&0.188&0.21& 861& 967& 571& 545& 537& 522& 4.30\\
21&30&35.22&-45& 4&35.8&0.340&0.47& 790&   0& 538& 595& 654&   0& 0.53\\
21&30&38.56&-45&17&43.7&0.110&0.31& 790&1225& 493& 874& 530& 425& 3.69\\
21&31& 4.79&-43&56&42.3&0.210&0.24&1010&   0& 621& 703& 648&   0& 3.59\\
21&31&15.03&-43& 3&32.7&0.196&0.98& 929&   0&3426& 955& 821&   0& 0.75\\
21&31&21.09&-45&42& 3.7&0.104&0.24&1037& 975& 589& 563& 594& 446& 4.25\\
21&31&15.90&-42&43&18.9&0.365&1.18&1306&1223& 983&1143&1095&   0& 7.43\\
21&31&21.75&-42&57&23.2&0.199&0.82& 984&   0& 535& 621& 721& 770& 4.16\\
21&31&44.75&-43&45&49.0&0.160&0.34& 836&2303& 589& 630& 546& 269& 2.99\\
21&31&50.70&-42&49& 0.8&0.182&0.20& 885&1390& 647& 543& 652& 427& 3.25\\
21&32& 6.07&-43& 5&25.8&0.283&0.25&1099&   0& 796& 929& 902&   0& 7.84\\
21&32&25.45&-45&53&15.0&0.104&0.15& 862&1169& 628& 505& 657& 431& 3.31\\
21&32&26.06&-44&59& 2.6&0.190&0.43&1135&   0& 630&1295& 681& 434& 3.83\\
21&32&44.01&-45&54&16.4&0.388&0.39& 800&   0& 578& 629& 609&   0& 1.96\\
21&32&51.48&-46&41& 5.8&0.353&0.89&1034& 767& 625& 795& 772&   0&25.42\\
21&33& 2.60&-44&16&33.7&0.325&0.28& 790&   0& 553& 492& 618&   0& 2.37\\
21&33&18.30&-45& 2&33.4&0.047&0.23& 883&   0& 544& 645& 688& 597& 4.75\\
21&33&15.03&-43&30&29.5&0.176&0.14& 827&1304& 586& 636& 590& 486& 4.04\\
21&34& 0.20&-45&27& 9.4&0.136&0.96& 691&   0&5772& 717& 690&3517& 0.58\\
21&34&48.23&-44&28& 8.1&0.494&0.52& 888&   0& 613& 919& 969&   0& 2.52\\
21&35& 0.77&-44&16&28.6&0.284&0.22& 951&1093& 591& 506& 584&   0& 2.15\\
21&35&53.01&-44&21&44.3&0.407&0.76&1228&1032&1498& 825& 884&   0& 0.91\\
21&35&54.17&-44&14&32.4&0.411&0.19& 970& 760& 515& 714& 629&   0& 5.15\\
21&36& 3.46&-45&14&35.3&0.103&0.29& 874&   0& 491& 559& 572& 518& 6.52\\
21&36&31.16&-44&20&53.9&0.432&0.26& 984&   0& 546& 693& 731&   0& 1.52\\
\end{tabular}
\end{table*}

\begin{table*}
\begin{tabular}{rrrrrrrrrrrrrrr}
\multicolumn{3}{c}{RA (1950)}&\multicolumn{3}{c}{Dec (1950)}&
 z&$\delta B$&[O II]&[Ne III]&H$\beta$&[O III]&[O III]&H$\alpha$&O/H\\
&&&&&&&&&&&&&&\\
21&36&49.17&-44&20&47.4&0.274&0.35& 870& 914& 602& 623& 616& 369& 3.37\\
21&37&17.79&-44&21&10.6&0.346&1.63&1022& 690& 637& 929& 741&   0& 3.74\\
21&37&24.65&-44&27&28.9&0.108&0.51& 796&   0& 465& 510& 533& 442& 1.82\\
21&37&56.97&-44&45&33.4&0.269&0.26& 994&3513& 668& 878& 699&   0& 2.95\\
21&37&57.13&-44&44&29.1&0.356&0.61& 832&1523&3048& 665& 789&   0& 1.18\\
21&37&56.15&-44&34&26.2&0.141&0.30& 951&   0& 620& 596& 639& 438& 3.39\\
21&38&19.28&-44&40&34.3&0.158&0.40& 938&   0& 484& 368& 355&   0& 0.78\\
21&38&32.36&-44&26&47.6&0.101&0.20&1082&   0& 577& 641& 636& 468& 3.63\\
21&39& 6.77&-44& 6&53.2&0.181&0.15& 908&1104& 642& 641& 668& 457& 4.24\\
\end{tabular}
\end{table*}

The main purpose of the spectroscopic follow-up was to confirm the
identity of the candidates as AGN, and obtain redshifts.  Given the
high surface density of candidates, the ideal intstrument for this
purpose was the 2dF multi-fibre spectrograph on the Anglo Australian
Telescope at Siding Spring.  In a 4 night run in July 2002, 1400 AGN
candidates were observed in 5 separate pointings making up an area of
about 12 deg$^{2}$ and including some overlap.  All spectra were
reduced using standard 2dF software to provide instrumental and
wavelength calibration, and also sky subtraction.  The feasibility
of flux calibrating 2dF data is somewhat controversial, and has
presented problems to other groups (\cite{c02}).  However, in the
region of interest of the spectra discussed here, the response function
is essentially flat, and the sensitivity only starts to decline
shortward of about 4500 $\AA$.  This is well clear of any redshifted
H$\beta$ lines which are the principal features of interest in this
paper.  On this basis it was decided that to introduce a controversial
procedure in a situation where the results of the analysis would not be
affected by its inclusion would be to add an unnecessary complication
to the line of argument.  Consequently, all spectra illustrated in this
paper are shown without flux calibration. 

Redshifts for all emission line objects were measured with the FIGARO
package.  Of the 894 candidates with successfully measured redshifts,
129 were found to have emission line spectra with redshift $z < 0.5$,
and these were used to form a catalogue of candidate Seyfert galaxies.
To determine a more precise classification for these low redshift
objects, the spectra were re-analysed with a view to measuring
line widths and line ratios.  To do this, the Gaussian fitting routine
in FIGARO was used.  This routine gives as outputs the fitted line
centre, full width at half maximum (FWHM) and line integral.  Although
the routine outputs error estimates for each measurement based
primarily on photon statistics, these were checked by comparing
duplicate measures of the same object and of skylines.  The calculated
errors from the FIGARO package were found to be too low, and so the
empirically determined errors were used.  These amounted to
approximately 0.6 $\AA$ or 30 km sec$^{-1}$.  The spectral resolution
as measured from the sky lines was 10.4 $\AA$ or 520 km sec$^{-1}$ at
6000 $\AA$.

The measurement of line integrals for the purposes of obtaining line
ratios is well known to be problematic as a number of effects, both
cosmological and instrumental, can cause problems.  In the case of
2dF data the instrumental effects are especially troublesome, as
accurate sky subtraction is hard to achieve, and flux calibration is
unreliable.  The normal way around these problems is to measure
line ratios of lines of similar wavelength when systematic effects
largely cancel out, and this is the approach we adopt in this paper.
The line ratio of principal interest here is
[O III]\,$\lambda 5007/$H$\beta$ where the error determined from
repeat measures of the same object was $\pm 5\%$.

Of the 129 objects with $z < 0.5$, 37 were Seyfert 1 galaxies 
with broad Balmer lines and relatively narrow forbidden lines.
The spectra of two of these are illustrated in Fig. 1, where the
width of the permitted lines and the [O III]\,$\lambda 5007/$H$\beta$
ratio put them firmly in the Seyfert 1 domain.  Among these broad line
objects there were also 5 narrow-line Seyfert 1 galaxies in the sense
of Osterbrock \& Pogge (1985) with H$\beta$
FWHM $\stackrel{<}{\sim} 2000$ km sec$^{-1}$ and
[O III]\,$\lambda 5007$/H$\beta \stackrel{<}{\sim} 3$ with S/N good
enough to be confident of their classification. The spectra of two of
these are shown in Fig. 2 (one with LINER characteristics), where it
will be seen that the Balmer lines are weak and narrow compared with
those in Fig. 1.

Spectra of the remaining 92 objects were characteristic of Seyfert 2 or
starburst galaxies with narrow forbidden and permitted lines of similar
width.  From these, 55 were selected as having adequate signal-to-noise
for further study.  In all cases, narrow [O III]\,$\lambda 5007$,
[O II]\,$\lambda 3727$ and other forbidden lines were clearly
seen, and H$\beta$ was narrow and weak, sometimes narrower
than the forbidden lines.  H$\alpha$ was similarly narrow where
observed in the lowest redshift objects.  Two typical examples are
illustrated in Fig. 3, and show all the normal characteristics of
Seyfert 2 galaxies.  The forbidden lines include
[O III]\,$\lambda 5007$, [O III]\,$\lambda 4959$,
[O II]\,$\lambda 3727$ [Ne III]\,$\lambda 3869$,
[SII]\,$\lambda 6716 + 6731$ and [OI]\,$\lambda 6300$ as well as narrow
Balmer lines and HeI\,$\lambda 5876$.  On the basis of the spectral
classification scheme of Veilleux \& Osterbrock (1987) these objects
lie in or near the region occupied by AGN, and are probably Seyfert 2
galaxies.

The measurement of line widths and other information for the sample of
emission line galaxies is given in Table 1.  The first six columns give
the Right Ascension and Declination for the sample members, and the
seventh column the redshift measured from the forbidden lines.
Column 8 gives the amplitude in the $B_{J}$ band measured over the
period from 1975 to 2002 and is described further below.  Columns 9 to
14 give the FWHM line widths in km sec$^{-1}$ of the
[O II]\,$\lambda 3727$, [Ne III]\,$\lambda 3869$,
H$\beta$\,$\lambda 4861$, [O III]\,$\lambda 4959$,
[O III]\,$\lambda 5007$ and H$\alpha$\,$\lambda 6563$ lines, measured
as described above.  No attempt has been made to correct the line
widths for instrumental broadening, and given the estimated spectral
resolution from the night sky lines, many of the forbidden lines are
effectively unresolved.  In cases where the line is redshifted out of
the observed passband, or is not seen in the spectrum a value of zero
is given.  The final column is the [O III]\,$\lambda 5007$/H$\beta$
line strength ratio which is commonly used as a discriminant between
various classes of active galaxies.  There is for the most part a good
correlation between the [O III]\,$\lambda 5007$ and
[O III]\,$\lambda 4959$ line widths, giving additional support to the
error estimates discussed above.  The [O II]\,$\lambda 3727$ line
tends to be of similar or greater width than the [O III] lines 

\subsection{Light curves}

\begin{figure*}
\begin{picture} (200,585) (10,200)
\includegraphics{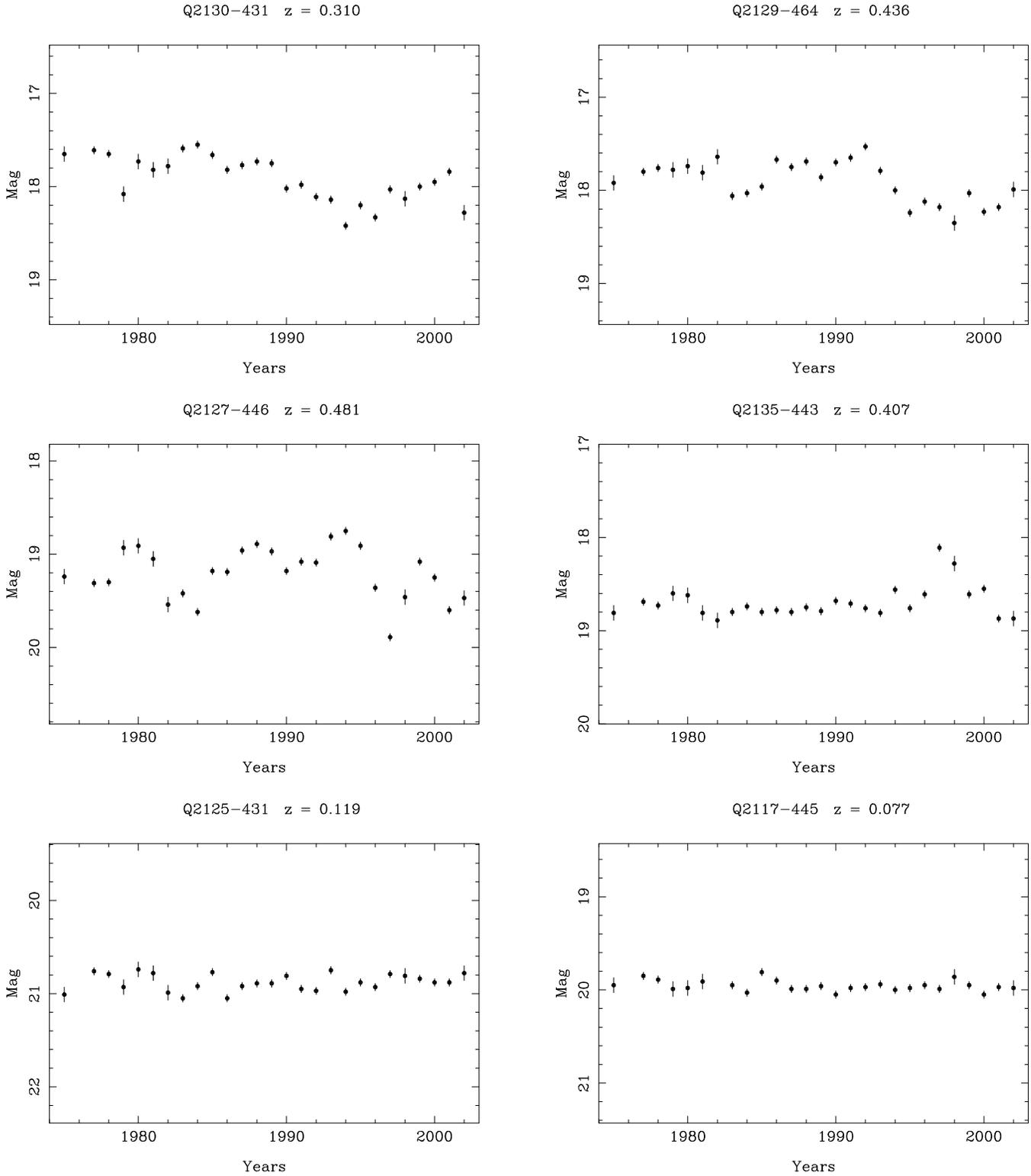}
\end{picture}
\caption{Light curves for Seyfert galaxies in the $B_{J}$ passband.
 The top two panels are for Seyfert 1 galaxies, the middle two for
 narrow line Seyfert 1 and the bottom two for Seyfert 2 or starburst
 galaxies.
 \label{fig4}}
\end{figure*}

The quasar monitoring programme in field 287 was started in 1975, and
between 1977 and 2002 at leat one and usually 4 plates in the $B_{J}$
passband (IIIa-J emulsion though a GG395 filter) have been taken each
year (see Hawkins (2003) for details).  The primary purpose of these
data is for Fourier analysis of the spectrum of variations, but they
are also well-suited to identifying morpological features in the
light curves, and measuring basic properties of the variation such as
amplitude. 

Fig. 4 shows the light curves of the of the objects with spectra in
Figs 1-3.  The top two panels are for Seyfert 1 galaxies, and they
show variations typical for this class of object.  The middle two
panels are for narrow line Seyfert 1 galaxies, and they too show
variations.  For the left hand panel the light curve is similar to the
normal Seyfert 1 galaxies, but for the right hand panel the variations
are smaller and the object hardly varies significantly for most of the
monitoring period.  The bottom two panels show light curves for the
objects classified from their spectra in Fig. 3 as Seyfert 2 or
possibly starburst galaxies.  It will be seen that they do not vary in
brightness within the observational errors which is in line with the
current understanding of the Seyfert 2 phenomenon in which the active
nucleus is being viewed as weak reflected polarised light which due to
its diffuse nature can only vary in brightnes on very long timescales.

\section{Naked active galactic nuclei}

\begin{figure*}
\setlength{\unitlength}{1mm}
\begin{picture} (200,45) (0,45) 
\includegraphics[width=0.5\textwidth]{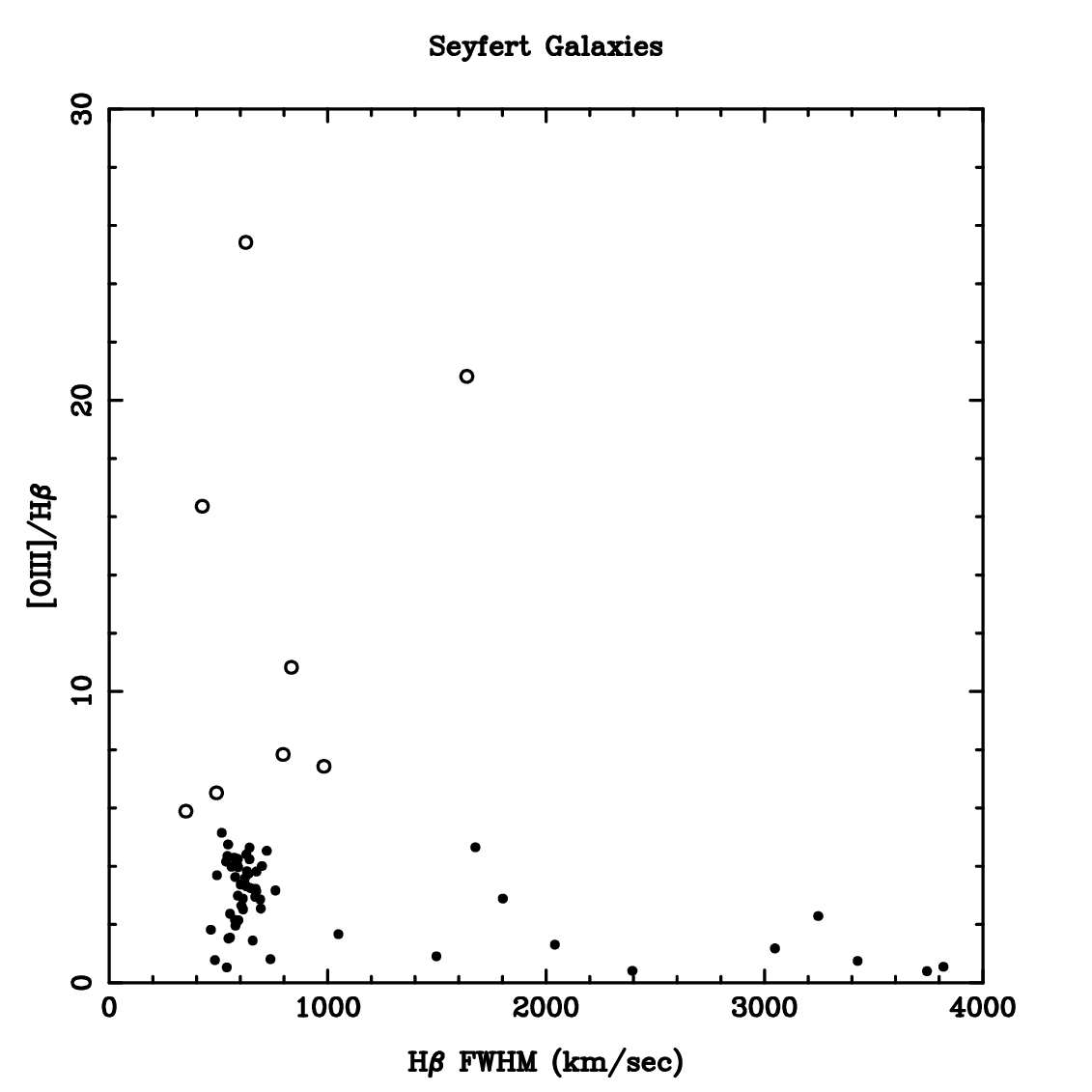}
\end{picture}
\begin{picture} (200,45) (-90,0) 
\includegraphics[width=0.5\textwidth]{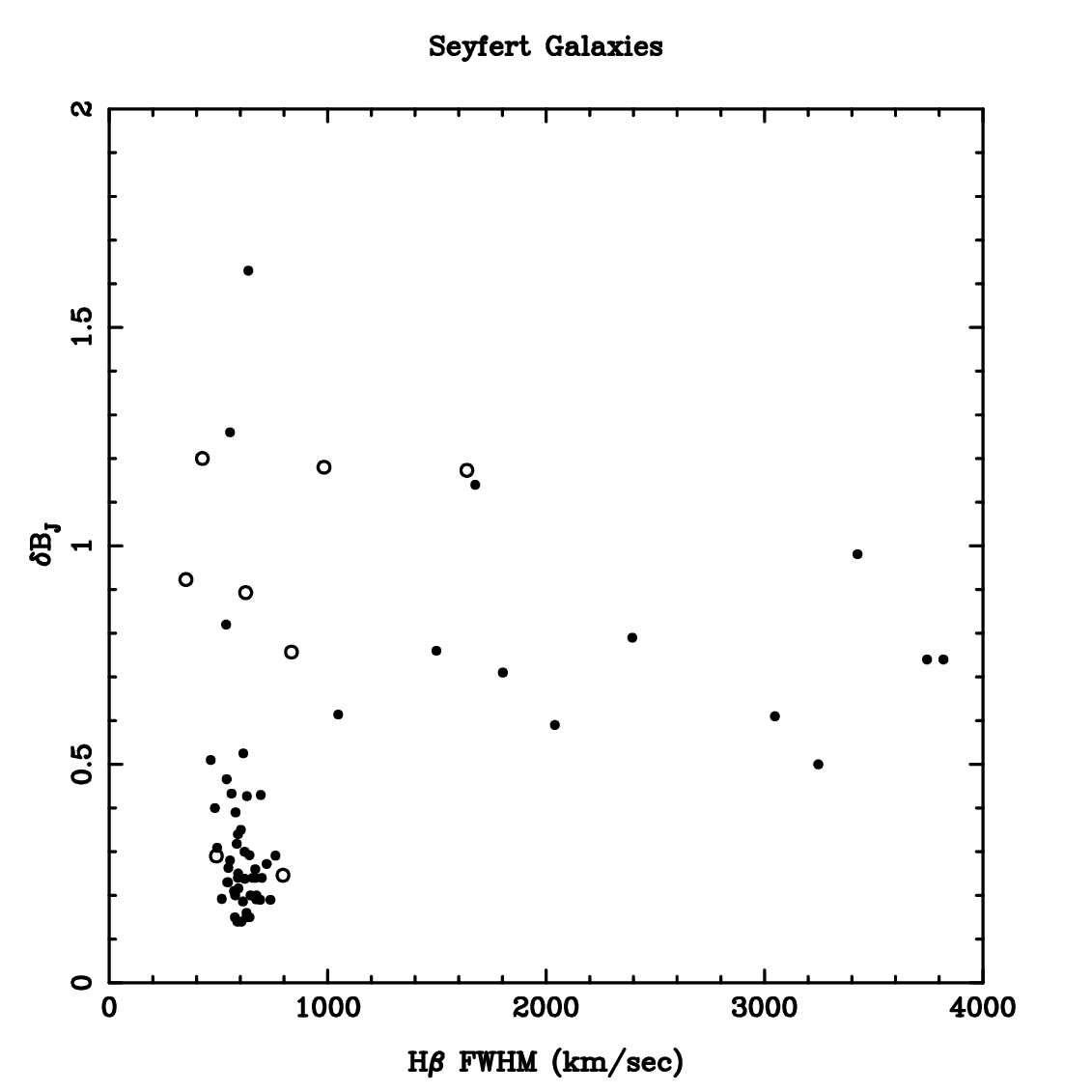}
\end{picture}
\caption{Relation between H$\beta$ line width and [O III]/H$\beta$ line
 strength ratio (left hand panel) and between H$\beta$ line width and
 amplitude (right hand panel).  Ojects with the largest
 [O III]/H$\beta$ ratio are shown as open circles.
 \label{fig5}}
\end{figure*}

Perhaps the most straightforward discriminant between the various
classes of AGN is the width of the Balmer lines.  For normal quasars
and Seyfert 1 galaxies the FWHM of these lines is typically 6000  -
10000 km sec$^{-1}$.  In the class of narrow line Seyfert 1 galaxies
(NLS) defined by Osterbrock \& Pogge (1985) it is usually less than
2000 km sec$^{-1}$, and often only slightly larger than the forbidden
line widths.  Another important defining characteristic of Seyfert
galaxies is the [O III]\,$\lambda 5007$/H$\beta$ line strength ratio
which is used to distinguish between Seyfert 1 and Seyfert 2 galaxies
(\cite{s81}).  In this paper they define Seyfert 2 galaxies as having
[O III]\,$\lambda 5007$/H$\beta > 3$, a definition which has stood the
test of time in subsequent studies (\cite{r00}).  The
[O III]\,$\lambda 5007$/H$\beta$ has also proved useful in
distinguishing between Seyfert 2 galaxies and H II region-like or
starburst galaxies, where a high values favours a Seyfert 2
classification (\cite{v87}).

  In Fig. 5 (left hand panel) we plot the relationship between H$\beta$
line width and [O III]\,$\lambda 5007$/H$\beta$ line strength ratio.
The diagram may be divided into three zones.  Firstly, Seyfert 1 type
galaxies or quasars may be seen along the bottom with
[O III]\,$\lambda 5007$/H$\beta < 3$ in all but one case, and a broad
Balmer line (H$\beta$ FWHM $> 1000$ km sec$^{-1}$.  In the bottom
left hand corner there is a compact group of objects with small
[O III]\,$\lambda 5007$/H$\beta$ and small H$\beta$ FWHM which is
the area occupied by starburst and some Seyfert 2 galaxies.  However,
for the most part Seyfert 2 galaxies have larger
[O III]\,$\lambda 5007$/H$\beta$ (\cite{v87}) and such objects may be
seen in the left hand side of the diagram.  These galaxies form a
well defined group in Fig. 5, and have been plotted as open circles.
On the basis of their large [O III]\,$\lambda 5007$/H$\beta$ ratio
and small H$\beta$ FWHM these objects are most likely to be Seyfert 2
galaxies, or possibly starburst galaxies.  Either way they should
definitely not vary in optical passbands.

The right hand panel of Fig. 5 shows the same objects,  but this time
with amplitude plotted as a function of H$\beta$ line width.  The
amplitude is taken from the monitoring programme described above, and
is the difference between maximum and minimum light in magnitudes in
the $B_{J}$ passband over a period of 27 years.  In this diagram, the
objects with large H$\beta$ line width identified as quasars all appear
with amplitudes of 0.5 magnitudes or more, as expected for quasars
(\cite{h00}).  Most of the objects in the bottom left of the left hand
panel classified as Seyfert 2 or starburst galaxies show little or no
significant variation, but six of the eight obects with large 
[O III]\,$\lambda 5007$/H$\beta$ shown as open circles have the
large amplitude normally only found for quasars or Seyfert 1 galaxies.
In fact in the right hand panel of Fig. 5 they seem to form a natural
extension of the quasar locus into the small H$\beta$ line width
regime. 

\begin{figure*}
\setlength{\unitlength}{1mm}
\begin{picture} (200,235) (-10,0)
\includegraphics[width=0.9\textwidth]{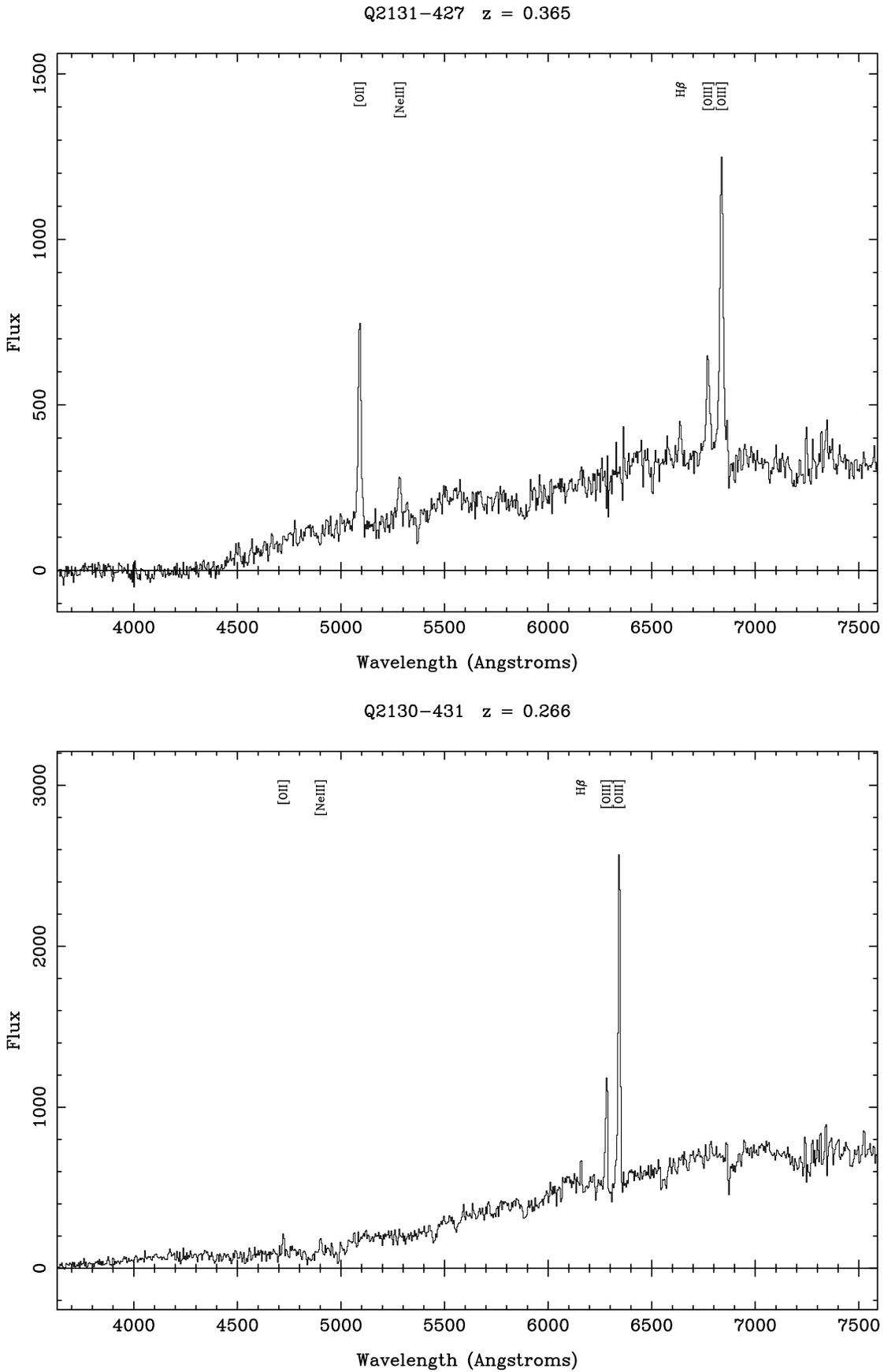}
\end{picture}
\caption{Spectra of variable Seyfert galaxies with Seyfert 2 type
  spectra (flux in arbitrary units).
 \label{fig6}}
\end{figure*}

\begin{figure*}
\setlength{\unitlength}{1mm}
\begin{picture} (200,235) (-10,0)
\includegraphics[width=0.9\textwidth]{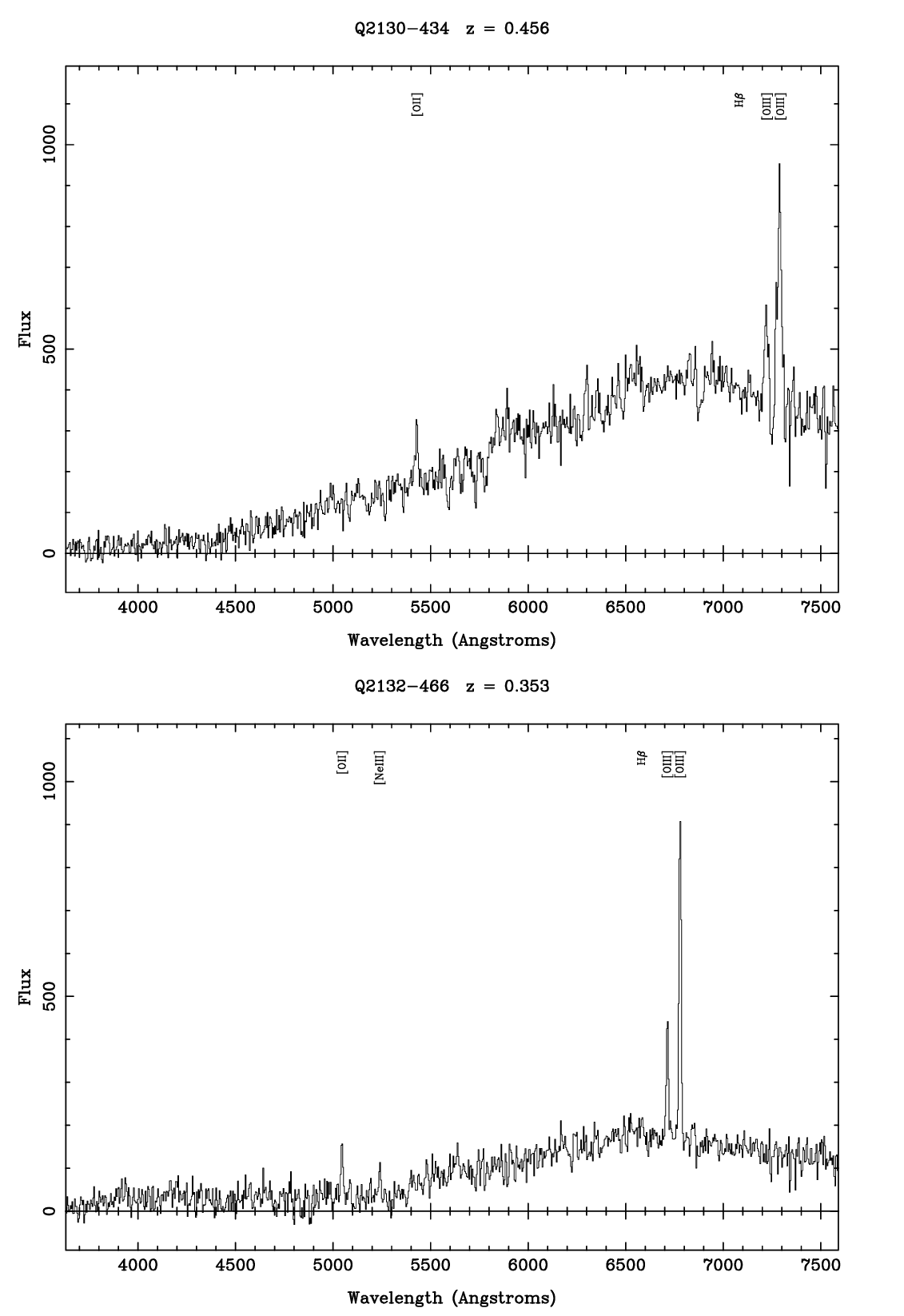}
\end{picture}
\caption{Spectra of variable Seyfert galaxies with Seyfert 2 type
  spectra (flux in arbitrary units).
 \label{fig7}}
\end{figure*}

\begin{figure*}
\begin{picture} (200,585) (10,200)
\includegraphics{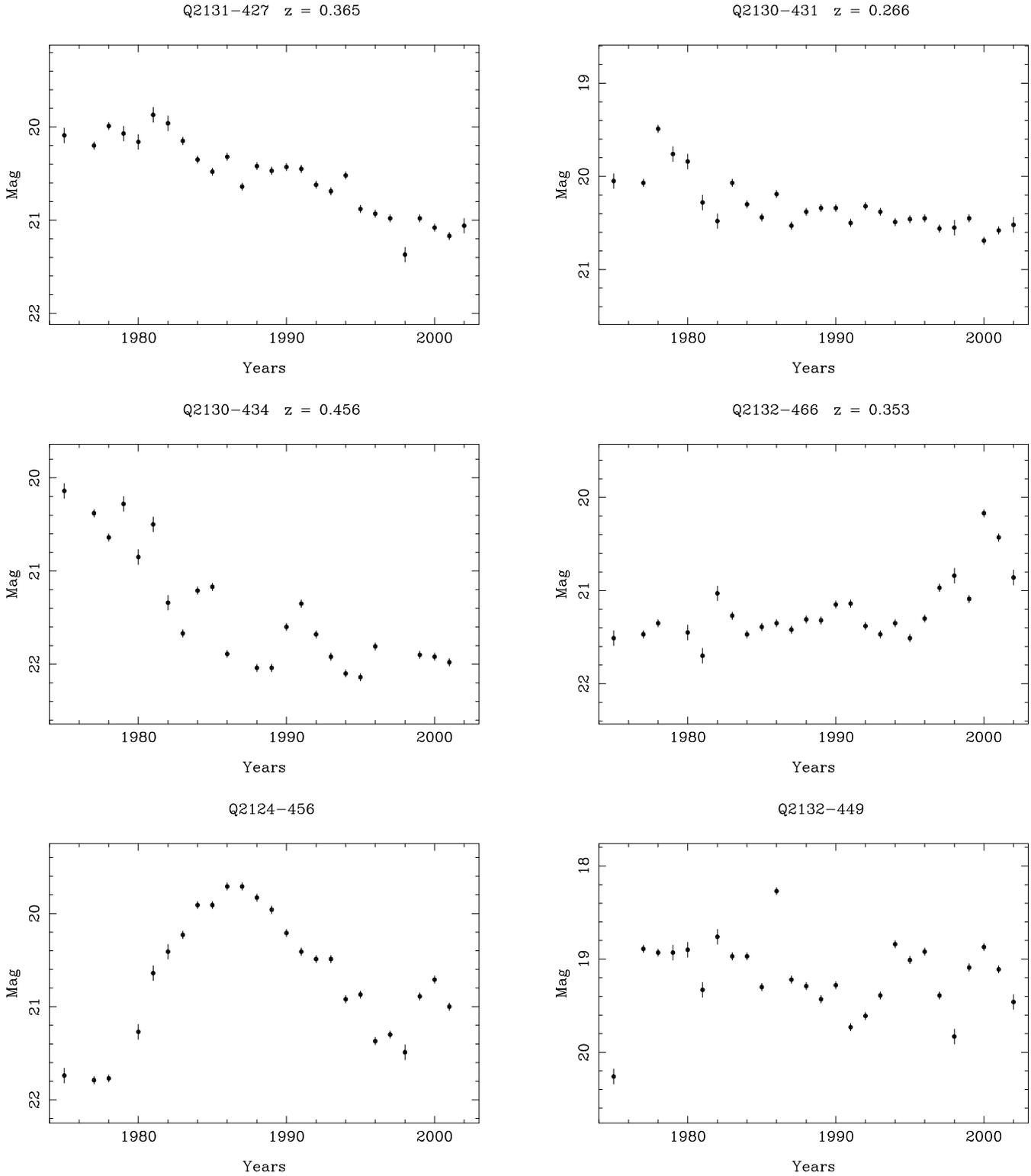}
\end{picture}
\caption{Light curves for active galaxies in the $B_{J}$ passband.
 The top four panels are for galaxies with Seyfert 2 type spectra,
 and the bottom two for galaxies with featureles spectra.
 \label{fig8}}
\end{figure*}

In Figs 6 and 7 we show the spectra of four of the large amplitude
objects shown as open circles in Fig. 5.  All four spectra show very
weak H$\beta$ of a similar or smaller FWHM to the forbidden lines,
and in each case there is no sign of a broad component to the line.
This is very characteristic of Seyfert 2 galaxies, although does not
absolutely exclude starburst galaxies (\cite{v87}).  The strength of
[SII]$(\lambda 6716 + \lambda 6731)$ or [OI]$\lambda 6300$ would
settle the matter, but unfortuneately in all four cases they are
redshifted out of the optical passband.  All four spectra show
strong narrow forbidden lines of [O III]\,$\lambda 5007$
[O III]\,$\lambda 4959$ and [O II]\,$\lambda 3727$, and in three
cases [Ne III]\,$\lambda 3869$.  There is good agreement between
the FWHM of the two [O III] lines, but the [O II] and [Ne III] lines
tend to be somewhat broader.  The H$\beta$ line on the other hand
is very weak and tends to be narrower than the forbidden lines. 
Other Balmer lines and the Mg II\,$\lambda 2798$ line are not detected.

On the face of it, the four objects in Figs 6 and 7 have spectra
indistinguishable from Seyfert 2 galaxies.  However their place among
the six large amplitude objects shown as open circles in the right hand
panel of Fig. 5 would appear to rule out this classification, or
indeed that of starburst galaxy.  To illustrate the nature of their
variability better, their light curves are shown in the top four panels
of Fig. 8.  It will be seen that all four objects show brightness
variations by a factor of 3 or more on a timescale of 5 to 10 years as
well as shorter term variations.This type of variation implies a direct
view of the compact nucleus itself.  Such strong rapid variability
could not be seen in the weak reflected light which is supposed to
characterise a Seyfert 2 nucleus.  Furthermore, the strong variations
in nuclear light can be used to put a lower limit on the luminosity of
the nucleus.  On the basis that the change in flux can only be due to
the nuclear light, then at maximum the absolute magnitudes of the
nuclei turn out to be $M_{B} \leq -21.5$.  This figure, which by the
nature of the derivation excludes any contribution from the underlying
galaxy, is typical for Seyfert 1 galaxies.  However, it is very much
greater than for Seyfert 2 galaxies where for example in the
well-studied (\cite{c00}) and archetypal NGC 1068 the continuum
emission amounts to a luminosity of $M_{B} \approx -15.3$.

For objects with luminous variable nuclei, the unified model of AGN
would lead one to expect a spectrum dominated by broad emission lines.
In particular, at this redshift one would expect to see strong broad
H$\beta$, H$\gamma$, H$\delta$ and Mg II emission lines as seen for
example in Figs 1 and 2.  As discussed above, the spectra shown in Figs
6 and 7 have all the characeristics of Seyfert 2 spectra.  The H$\beta$
lines are weak and narrow (see Table 1), and show no sign of a broad
component.  There are several forbidden lines in the spectra, all of a
similar width or broader than the H$\beta$ line, and the
[O III]\,$\lambda 5007/$H$\beta$ ratios put them in the domain of
Seyfert 2 galaxies (\cite{v87}), and certainly outside the parameter
space occupied even by narrow line Seyfert 1 galaxies (\cite{o85}).  
However, the luminosity and variability of the nuclei make it clear
that we are not viewing a Seyfert 2 galaxy.  The only plausible
explanation appears to be that these are AGN without a broad line
region.

\section{Discussion}

\begin{figure*}
\setlength{\unitlength}{1mm}
\begin{picture} (200,235) (-10,0)
\includegraphics[width=0.9\textwidth]{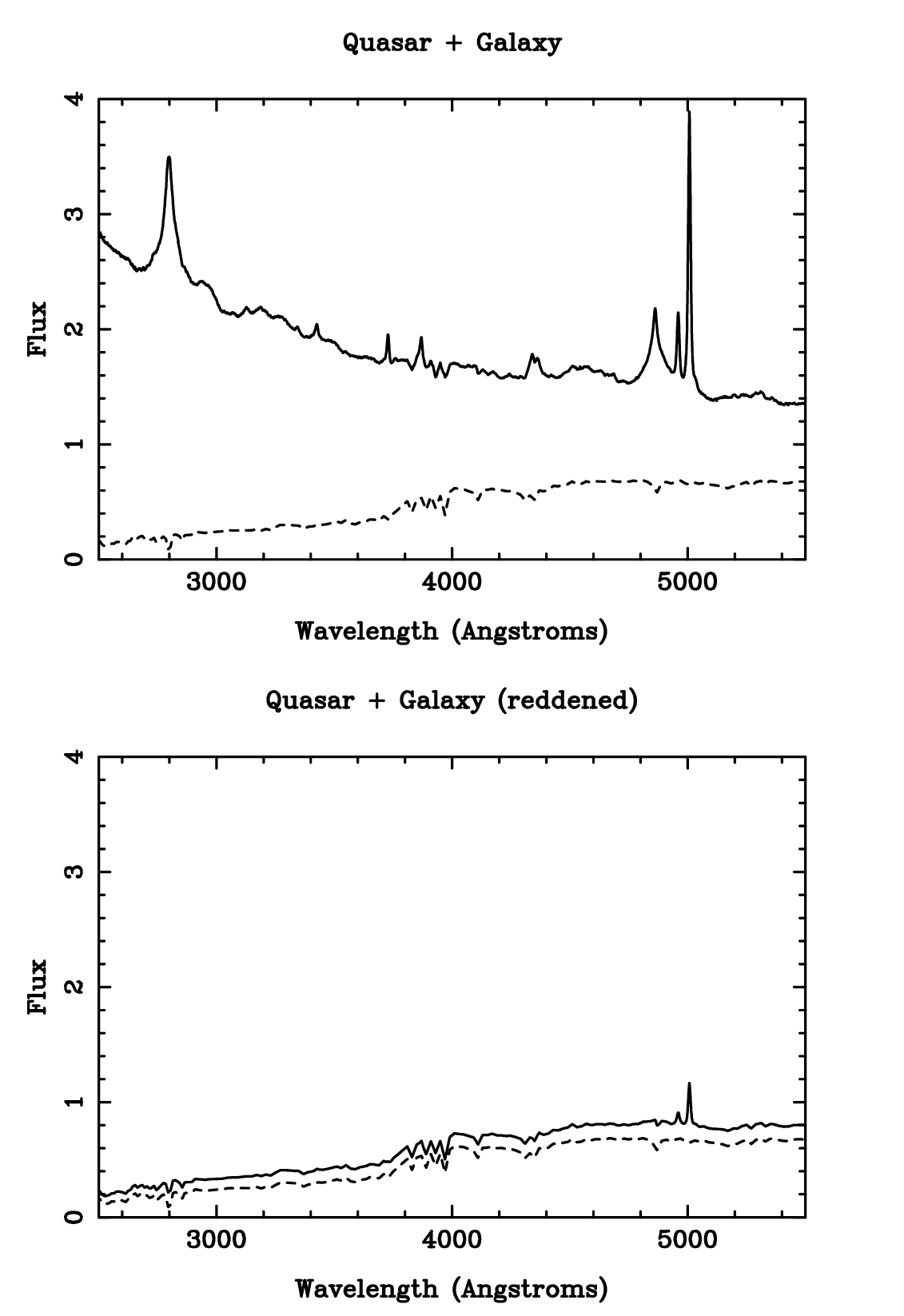}
\end{picture}
\caption{Simulated spectra of quasar plus host galaxy (solid line) and
 galaxy alone (dashed line).  The top panel shows the quasar spectrum
 unreddened, and the bottom panel with reddening sufficient to make the
 broad H$\beta$ line undetectable. Flux in arbitrary units.
 \label{fig9}}
\end{figure*}

\begin{figure*}
\setlength{\unitlength}{1mm}
\begin{picture} (200,117) (-10,0)
\includegraphics[width=0.9\textwidth]{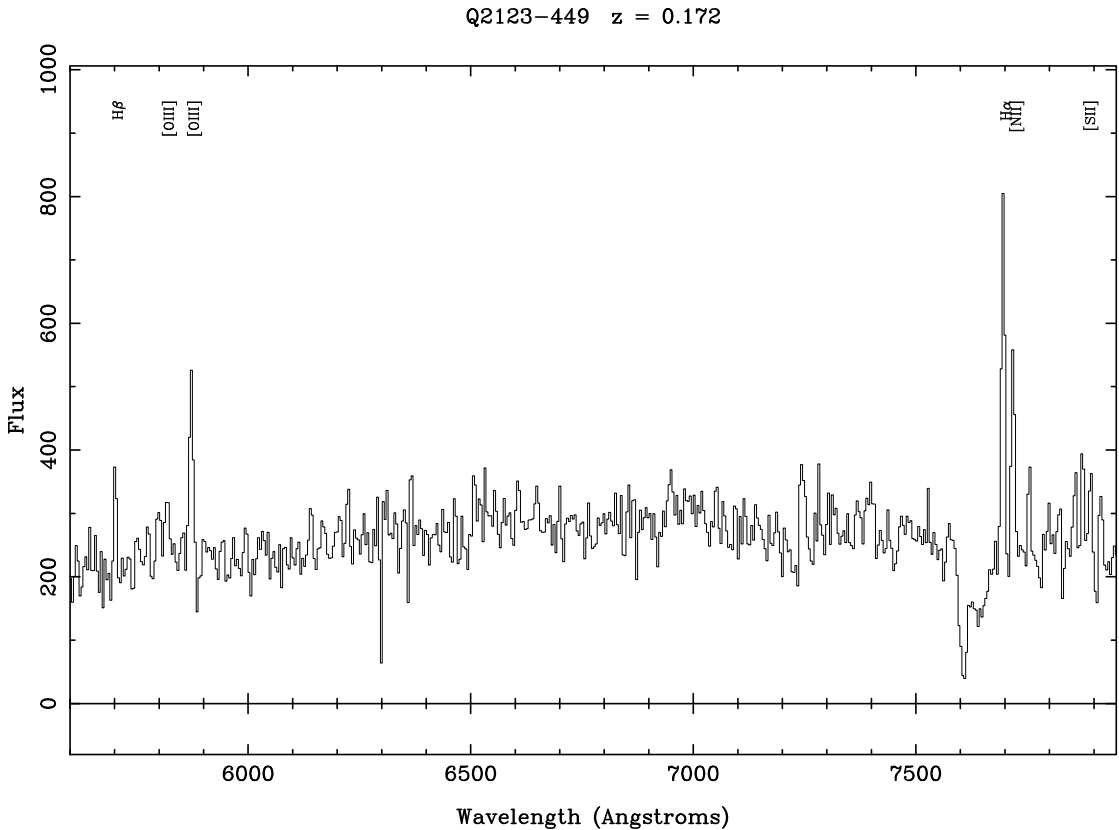}
\end{picture}
\caption{Spectrum of a variable Seyfert galaxy showing Seyfert 2 type
 Balmer lines (flux in arbitrary units).
 \label{fig10}}
\end{figure*}

One possible explanation for the absence of broad emission lines in 
AGN spectra is that they fade in response to decrease in the continuum
flux.  Although it is hard to rule this possibility out, it can be
checked for Q2131-427 for which an earlier spectrum exists.  This
object was observed in 1991 with EFOSC on the 3.6m at ESO as part of
an earlier follow-up to the monitoring programme.  It will be seen
from Fig. 8 that at that time it was 0.7 magnitudes brighter than in
2002 when the spectrum in Fig. 6 was obtained.  The early spectrum is
however essentially unchanged apart from additional flux at the blue
end of the continuum.  There is still no sign of any broad emission
lines, and the H$\beta$ line is narrow and weak as in Fig. 6.

Another possibility is that the spectra in Figs 6 and 7 show Seyfert
1.8 or 1.9 galaxies in a low state.  These galaxies, first described by
Osterbrock (1981), have a weak or undetectable broad component to the
H$\beta$ line but broad H$\alpha$ emission.  These galaxies are known
to vary between this state and a more normal Seyfert 1 type spectrum,
and a case has been made by Goodrich (1989) that this change is due to
variable reddening.  To test this mechanism for the objects in our
sample we have simulated the effect of reddening a Seyfert 1 type
spectrum to the point where the broad H$\beta$ line becomes washed out
in the spectrum of the underlying galaxy to determine whether the blue
continuum is still detectable. For a model quasar spectrum we used the
composite from the FIRST bright quasar survey (\cite{b00}), and for the
underlying galaxy the models of Jimenez et al. (2004).  The top panel of
Fig. 9 shows the combined quasar and galaxy spectrum as a solid line,
and the galaxy contribution as a dashed line.  In the bottom panel the
quasar spectrum has been reddened to the extent that the broad H$\beta$
line is no longer visible, which results in the continuum emission
being extinguished to the extent that it is small compared with the
galaxy light, and insufficient to cause significant variation of the
combined flux.

A feature of Seyfert 1.8 and 1.9 Seyfert galaxies is that even when the
broad H$\beta$ line is washed out, the broad H$\alpha$ line remains
relatively strong and clearly visible(\cite{o81}).  Although the
objects in Figs 6 and 7 are at too high a redshift to observe the
H$\alpha$ line in the optical, there is one variable AGN in the sample
at a sufficiently low redshift to test for a broad H$\alpha$ component.
This object was not included in the sample of naked AGN candidates as
the signal-to-noise of the spectrum is rather poor, but it is
sufficiently good to throw some light on the presence of a broad
H$\alpha$ line.  The spectrum is shown in Fig. 10, and it will be seen
that both the H$\alpha$ and H$\beta$ lines are narrow, of comparable
width to the forbidden lines, and in neither case is there any
indication of the presence of a broad component.

The broad line region has been seen as a fundamental part of the
structure of AGN, although the unified model explains its presence
rather than providing a compelling argument for its existence.  It is
now clear that in a significant fraction of AGN the broad line region
is not present. 
Although here has been much debate over the structure of the BLR, there
seems litle doubt that it involves an interaction between the central
accretion disc and a surrounding star cluster (\cite{w99}). The idea
is that gas from the cluster falls onto the accretion disc causing a
complex and unstable high velocity flow pattern in which the inflowing
material is eventually accreted onto the central balck hole or ejected
as bi-polar winds.  This high velocity turbulent gas is ionised by
radiation from the accretion disc, resulting in the broad line
emission.  Williams et al. (1999) make the case that given inflowing
material, this picture is virtually inevitable.  On this basis it would
thus appear that for those AGN where broad emission lines are not
present, the source of inflowing material has dried up.  These naked
AGN therefore would represent a period in the duty cycle intermediate
between activity and dormancy.  They would survive on the fuel reserves
present in the accretion disc.

Such `naked' AGN appear to be quite common, comprising roughly 10\% of
the sample of the emission line galaxies observed, the remainder being
Seyfert 1 galaxies with broad permitted lines, or non-variable objects
with Seyfert 2 type spectra, which are presumably normal Seyfert 2
galaxies or starburst galaxies.

An interesting question which arises concerns the possibility that
naked nuclei may be found among more luminous AGN in the quasar
regime.  This almost inevitably means that candidate objects will
be at higher redshift, and much more difficult to identify without
the help of a system of strong narrow forbidden lines.  There is
much anecdotal evidence from surveys based on UVX selection for
objects with featureless spectra for which there is no plausible
classification.  Such objects are typically ignored for lack of
observational evidence as to their nature.  With information on
variability, one can say a lot more.  The bottom two panels of Fig. 8
show the light curves for two highly variable objects with variations
characteristic of quasars (\cite{h96,h02}).  However, their spectra,
shown in Fig. 11, are featureless.  Although this is not conclusive, it
is highly suggestive that we are seeing a more luminous AGN or quasar
without a broad line region.  It is possible that these are BL Lac
objects, but such objects tend to vary more violently and on shorter
timescales.  Further observations including a better spectrum and
and polarisation measures should clarify this, but for the moment it
must suffice to say that a quasar without a BLR is a plausible
classification. 

\begin{figure*}
\setlength{\unitlength}{1mm}
\begin{picture} (200,235) (-10,0)
\includegraphics[width=0.9\textwidth]{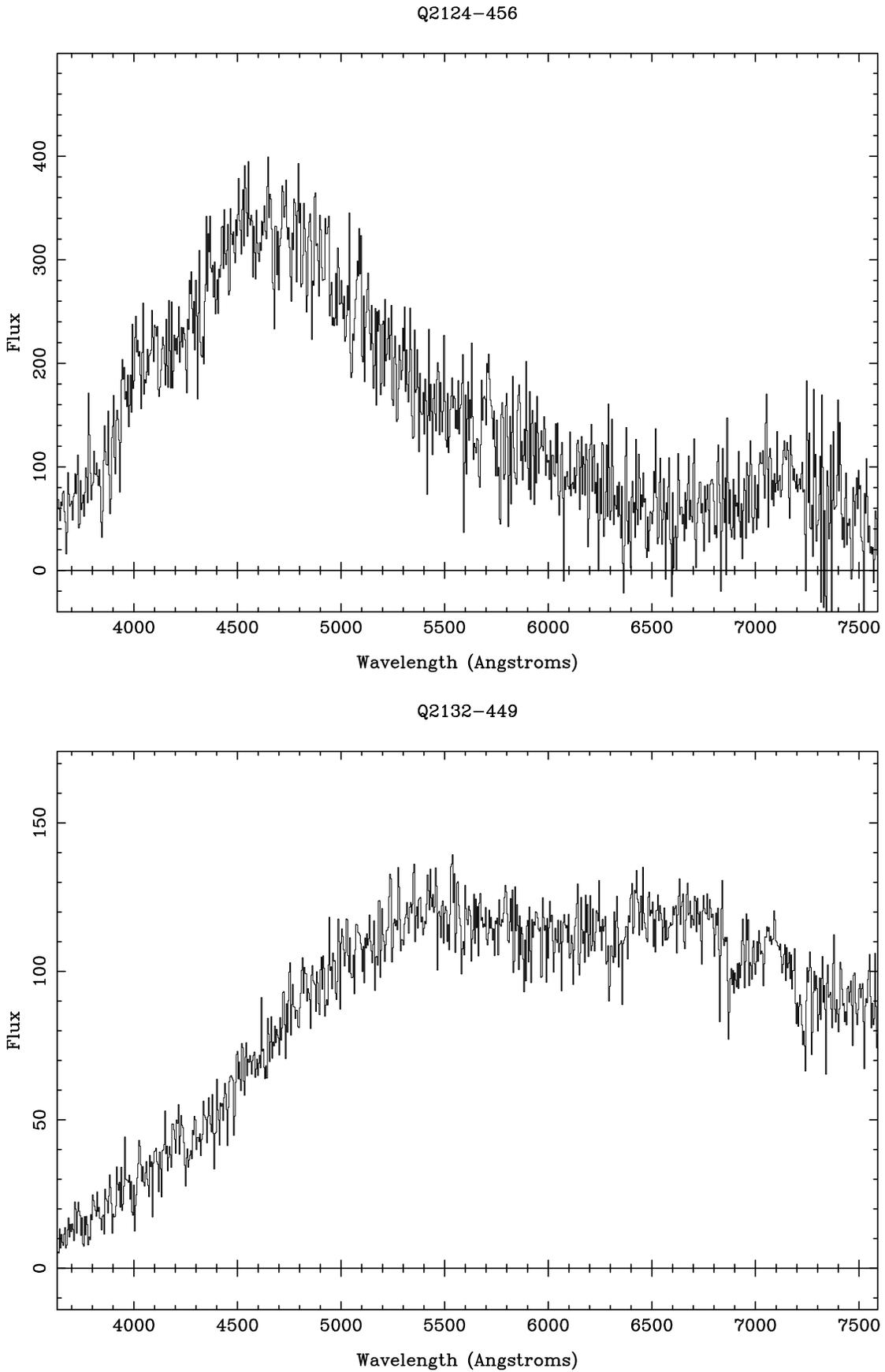}
\end{picture}
\caption{Spectra of variable objects with featureless spectra (flux in
 arbitrary units).
 \label{fig11}}
\end{figure*}

\section{Conclusions}

In this paper we have reported the discovery of a new class of AGN in
which the broad emission line region is completely absent although the
nucleus is viewed directly.  These naked AGN were discovered as part of
a survey and monitoring programme of AGN over a timescale of 25 years.
Their spectra show the narrow forbidden lines characteristic of Seyfert
2 galaxies, combined with very weak narrow Balmer lines with no
detectable broad component.  However, The
light curves show variations in brightness of at least a factor of
three in all cases implying bright variable continua.  We discuss other
examples from the literature of AGN with weak broad emission lines,
and investigate the possibility that the absence of a broad line
component could be due to reddening.  Our conclusion is thst we are
viewing the nucleus directly, and that in these AGN there is no broad
line region.  These objects comprise some 10\% of the sample initially
classified
as emission line galaxies and appear to be distinct from other
previously published Seyfert 1 galaxies with weak or variable emission
lines.  We also illustrate some possible examples of luminous AGN or
quasars with no broad emission lines.

These results are discussed in the context of current models for AGN,
and it is argued that the central star cluster fuelling the accretion
disc has ceased supplying gas, and the naked AGN represent a transition
stage between activity and a dormant phase, in which the disc draws on
its internal resources to maintain its energy output.

\end{document}